\documentclass[12pt,preprint]{aastex}

\usepackage{lscape}
\usepackage{wasysym}
\usepackage{natbib}
\usepackage{longtable}

\shortauthors{Fischer et al.}
\shorttitle{AGN Inclinations}

\begin{document}

\title{A Minor Merger Caught In The Act Of Fueling The AGN In Mrk 509}

\author{T.C. Fischer\altaffilmark{1},
D.M. Crenshaw\altaffilmark{1},
S.B. Kraemer\altaffilmark{2},
H.R. Schmitt\altaffilmark{3},
T. Storchi-Bergmann\altaffilmark{4},
R.A. Riffel\altaffilmark{5}}

\altaffiltext{1}{Department of Physics and Astronomy, Georgia State 
University, Astronomy Offices, 25 Park Place, Suite 600,
Atlanta, GA 30303; fischer@astro.gsu.edu}

\altaffiltext{2}{Institute for Astrophysics and Computational Sciences,
Department of Physics, The Catholic University of America, Washington, DC
20064}

\altaffiltext{3}{Naval Research Laboratory, Washington, DC 20375}

\altaffiltext{4}{Departamento de Astronomia, Universidade Federal do Rio Grande 
do Sul, IF, CP 15051, 91501-970 Porto Alegre, RS, Brazil}

\altaffiltext{5}{Departamento de F\'isica, Centro de Ci\^encias Naturais e Exatas, 
Universidade Federal de Santa Maria, 97105-900 Santa Maria, RS, Brazil}

\begin{abstract}

In recent observations by the {\it Hubble Space Telescope} ({\it HST}) as part of a 
campaign to discover locations and kinematics of AGN outflows, we found that Mrk 509 
contains a 3$''$ ($\sim$2100 {\it pc}) linear filament in its central region. Visible in 
both optical continuum and [OIII] imaging, this feature resembles a `check mark' 
of several knots of emission that travel northwest to southeast before jutting 
towards the nucleus from the southwest. Space Telescope Imaging 
Spectrograph (STIS/{\it HST}) observations along the inner portion of the filament reveal 
redshifted velocities, indicating that the filament is inflowing. We present further 
observations of the nucleus in Mrk 509 using the {\it Gemini} Near-Infrared 
Integral Field Spectrograph (NIFS), from which we conclude that this structure 
cannot be related to previously studied, typical NLR outflows and instead embodies 
the remains of an ongoing minor merger with a gas-rich dwarf galaxy, therefore providing a 
great opportunity to study the fueling of an AGN by a minor merger in progress. 

\end{abstract}

\keywords{galaxies: active, galaxies: Seyfert, galaxies: kinematics and dynamics, galaxies: individual(Mrk 509)}

~~~~~

\section{Introduction}
\label{sec1}

The fueling of active galactic nuclei (AGN) and their subsequent `feedback' 
via outflows of ionized gas is thought to play a critical role in the 
formation of large-scale structure in the early Universe \citep{Sca04}, chemical 
enrichment of the intergalactic medium \citep{Kha08}, and self-regulation of 
super-massive black hole (SMBH) and galactic bulge growth \citep{Hop05}. However, we have very little 
direct information on the details of AGN feeding and feedback. Large-scale 
fueling of AGN is thought to be instigated by galaxy interactions, and, in the 
case of luminous quasars, there is observational evidence that major mergers 
between large galaxies play a significant role \citep{Can01}. However, in the 
case of nearby, lower luminosity (L$_{bol}<10^{45}$ erg s$^{-1}$) Seyfert galaxies, 
the evidence is not so clear, as some studies claim that Seyferts show 
an excess of companions (e.g., \citealt{Kos11}), indicating large-scale interactions, 
while others claim a lack thereof (e.g., \citealt{Sch01b}). An obvious alternative to requiring 
{\it major} mergers is that lower-luminosity AGN are fueled by {\it minor} 
mergers of their host galaxies with small satellite galaxies. However, the 
observational evidence thus far has been lacking.

\begin{figure}
\centering
\includegraphics[width=0.45\textwidth]{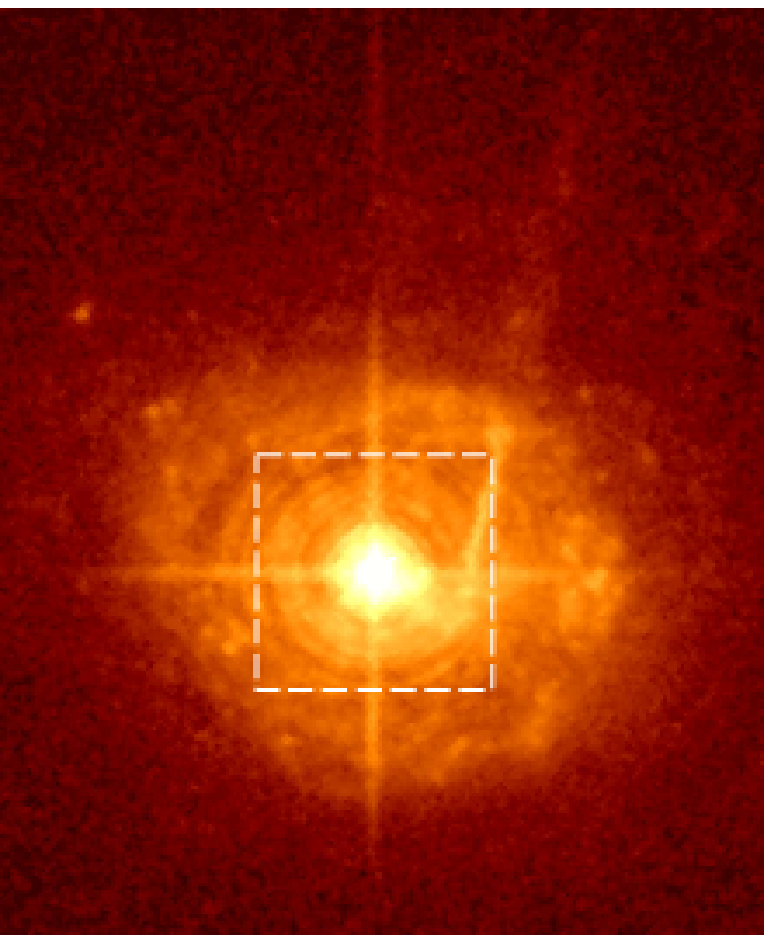} 
\caption[Mrk 509 NLR Image]{{\it HST} FQ508N narrow-band image of Mrk 509 showing primarily 
[O~III] emission. The filament can be seen to the right of the nucleus, extending from northwest to 
southeast before making a 90$^{\circ}$ turn toward the nucleus. Starburst activity 
can be seen in a ring around the nucleus at a radius of $\sim 3''$. The dashed box 
shows the $3'' \times 3''$ field of view observed with NIFS.}

\label{oiii}
\end{figure}

As part of a recent project to map the kinematics of the Narrow Line Region 
(NLR) of Seyfert galaxies \citep{Fis13}, we obtained WFC3 [OIII] and continuum 
(F550M) images of Mrk 509, a nearby ($z = 0.0346, 1'' \sim 700 {\it pc}$), highly 
luminous (L$_{bol} = 3.5 \times 10^{45}$ ergs s$^{-1}$, \citealt{Awa01,Kra04}) Type 
1 AGN, where we discovered an asymmetric filament adjacent to the nucleus, shown 
in Figure \ref{oiii}. The majority of the filament is a linear feature several 
kiloparsecs long to the west of the nucleus, stretching from northwest to 
southeast, and is most prominent at radii $< 2100$ {\it pc}, inside the 
massive starburst ring which surrounds the AGN (although it does extend beyond 
the ring; Figure \ref{oiii}). The linear feature connects to a second emission-line 
region extending from the southwest to the nucleus, giving the entire filament 
the appearance of a check mark.

In our {\it HST} kinematics analysis of a STIS long-slit spectrum of the inner 
portion of the filament \citep{Fis13}, we suggested that we may 
be viewing a minor merger with a dwarf galaxy and that this system would provide 
a great opportunity to study the fueling of an AGN by a minor merger in progress. 
However, from our initial results, we could not be certain, as the data could 
be interpreted in several ways, such that the observed redshifted velocities in the inner 
portion could be infalling from the near side of the galaxy or outflowing from the far 
side. Here we present further analysis of the filament using the {\it Gemini} NIFS 
Integral Field Unit (IFU), particularly of the extended linear feature, which clarify 
the kinematics of the filament and indicate that it is inflowing. 

\section{Observations}
\label{sec2}

We observed Mrk 509 using Gemini/NIFS employing the Gemini North Altitude Conjugate 
Adaptive Optics for the Infrared (ALTAIR) adaptive optics system in 2013 June 
through program GN-2013A-Q-40. We obtained observations in the Z-band, which 
has a spectral resolution of $R = 4990$ and covers a spectral region of 0.94 - 1.15 
$\mu$m. Observation sequencing followed standard object-sky-object dithering 
\citep{Rif11,Rif13} with off-source sky positions common for extended targets. Eight 
individual exposures of 450 s were obtained, for 3600 s in total. 
Data reduction was performed using tasks contained in the NIFS subpackage within the 
GEMINI IRAF package, in addition to standard IRAF tasks. The reduction process included 
image trimming, flat fielding, sky subtraction, and wavelength and s-distortion 
calibrations. Frames were corrected for telluric bands and flux calibrated by 
interpolating a blackbody function to the spectrum of a telluric standard star. The 
resultant data cube was median combined into a single data cube 
via the gemcombine task of the GEMINI IRAF package. The final data cube contains 
approximately 4000 spatial pixels (spaxels). At a redshift of $z = 0.03454$, each 
spaxel corresponds to an angular sampling of $0.05'' \times 0.05''$ (34.4 {\it pc} $\times$ 
34.4 {\it pc}), with observations covering the inner $3'' \times 3''$ (2.06 {\it kpc} $\times$ 
2.06 {\it kpc}) of the AGN. We note that the observations oversample the point-spread 
function, which has a full-width at half-maximum of 2 pixels.

\section{Analysis}
\label{sec3}

\begin{figure*}
\centering
\includegraphics[width=\textwidth]{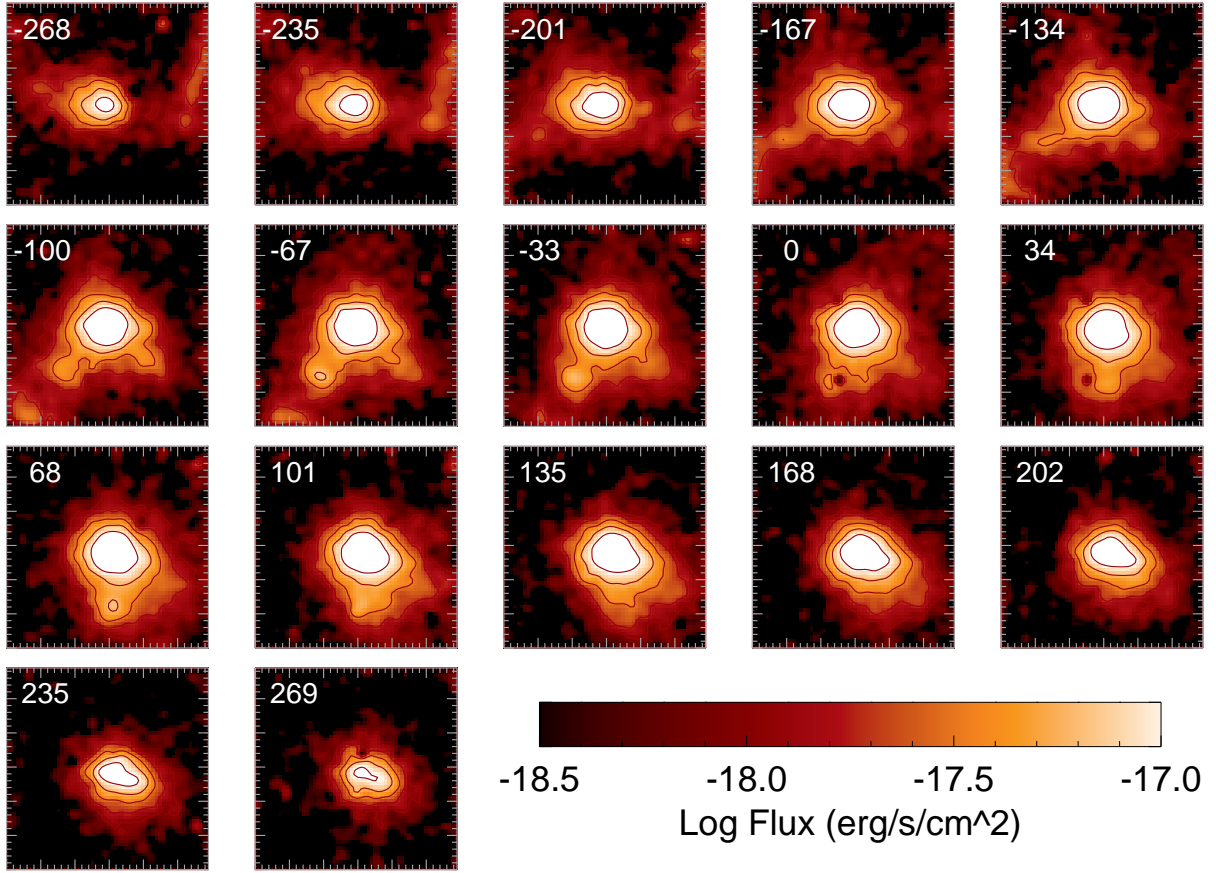}
\caption{$3'' \times 3''$ velocity channel maps of Mrk 509
obtained by integrating continuum-subtracted flux within velocity bins of 33 km s$^{-1}$ along the [S~III] emission-line 
profile. Numbers in the upper left corner of each map show the central velocity of each bin in km s$^{-1}$ from systemic velocity.}
\label{map}
\end{figure*}

Figure \ref{map} shows continuum subtracted [S~III] 0.95$\mu$m flux maps 
obtained by integrating the flux within velocity bins of 33 km s$^{-1}$ along the [S~III] 
emission-line profile.
We fit individual components of the [S~III] 0.95$\mu$m emission lines with Gaussians 
over an average continuum taken from line-free regions throughout the spectrum to 
measure velocities in each spaxel. A majority of the spaxels contained 
single-component [S~III] lines, which were fit with a single Gaussian (Figure 
\ref{ifu}, left), whereas the remaining spaxels contained 
two-component [S~III] emission lines, either as two individual peaked lines or a 
single peaked line with an asymmetric wing, which were fit with two Gaussians. 
The center and right panels of Figure \ref{ifu} display the kinematics of 
shorter wavelength (`blue') and longer wavelength (`red') components, respectively, 
for each [S~III] line fit with two Gaussians. If an emission line contained a peak 
and an asymmetric wing, the wing was fit with a Gaussian only in situations where 
the flux component responsible for creating a significant asymmetry 
in an emission line was traceable through adjacent spaxels in the cube.

The central wavelength of each Gaussian was used to measure a Doppler shifted 
velocity for each emission line component, given in the rest frame of the 
galaxy and using a [S~III] rest wavelength of 9533.2 \AA. We employ a Gaussian fit 
rather than a direct integration across the line profile because in most cases the 
former is more suited to extract individual velocities from blended lines. Noisy 
spectra (S/N $<$ 3 per resolution element and not adjacent to successfully fit 
lines) were not fitted.

The inner portion of the check-mark filament is visible in the velocity channel map to the 
right of the nucleus (Figure \ref{map}), approximately 0.5$''$ - 1.5$''$ west, with a outer, linear 
portion traveling from northwest to southeast before taking a sharp turn northeast 
approximately 0.25$''$ south of the nucleus. Velocities associated with the linear 
portion of the filament are clearly defined, with blueshifted velocities that 
increase with distance from the turn in the filament. The kinematics of the inner 
portion of the filament are more difficult to distinguish, as several spectra taken 
over the inner portion contain two component [S~III] lines. If we ignore all 
kinematics pertaining to the filament (i.e. the western blueshifted velocities 
of Figure \ref{ifu}), the remaining kinematics resemble disk rotation. This observation concurs with 
those made in \citet{Phi83}, who measured the kinematics of low-ionization 
lines surrounding the nucleus and found the velocities fit a disk model with a 
rotation axis of $PA = 135^{\circ}$. Adding the red component velocities measured 
in the double-peaked [S~III] lines to the map of the single line velocities creates 
a kinematic field that further resembles a rotating disk. 

In order to confirm that these kinematics could be due rotation of the host disk, 
We used {\tt DiskFit} \citep{Spe07,Sel10,Kuz12}, a publicly available code which 
fits non-parametric models to a given velocity field, in order to confirm that 
these kinematics could be due rotation of the host disk. We applied 
the rotation model to the kinematics, using the single line plus red component 
velocity field, an initial rotational major axis position angle of -135$^{\circ}$ 
and ellipticity of 0.18 (taken from \citealt{Phi83} kinematics and isophotes 
respectively), and a nuclear location centered in the NIFS field of view (FOV), near 
the [S~III] emission peak. For {\tt DiskFit} to create a kinematic model which has a 
center overlapping the photometric 
center, we required that the velocities at the photometric center be effectively zero. 
This required us to calculate a new redshift to replace the value of $z = 0.034397$ we had 
been using previously. [S~III] emission lines over the photometric nucleus were measured 
to have a central wavelength of 9863.5\AA~ and setting the velocities of these lines to zero, 
while also accounting for a heliocentric velocity of 31.5 km s$^{-1}$, we calculated a new 
redshift for Mrk 509 of $z = 0.03454$. All figures contain systemic velocities calculated 
using this new redshift. The model using these revised velocities generates a disk with a 
major axis $PA = -109.8^{\circ} \pm 0.2^{\circ}$, an ellipticity of $1-b/a = 0.05 \pm 0.01$, and 
an inclination of $i = 18.19^{\circ} \pm 1.06^{\circ}$, with a nucleus offset from the center of 
the FOV by $-0.28 \pm 0.05$ and $-1.0 \pm 0.11$ spaxels in the X and Y directions respectively. 
With a disk inclination of $18^{\circ}$, the maximum observed rotational velocities of $\sim$160 
km s$^{-1}$ can be deprojected to obtain a true rotational velocity of $\sim$ 500 km s$^{-1}$. 
While rather high, this velocity is not outside the realm of velocities measured in other 
disk galaxies \citep{Spa00}. Alternatively, rotation models are prone to showing a degeneracy 
between the amplitude of the rotation curve and the inclination of the galaxy, and the true 
velocity may be lower as the true inclination could be larger than what is given from the model 
output.

Figures \ref{rotation} and \ref{model} show the rotation and filament velocity field data
and the derived rotation model and residuals, respectively. The resultant major axis is practically 
identical to the isophotal major axis $PA = -112^{\circ}$ \citep{Phi83}, suggesting that these 
kinematics are indeed due to rotation. 
Large kinematics discrepancies exist approximately 0.75$''$ southwest of the 
nucleus, where residuals between the rotation data and model reach velocities of 
$\sim$100 km s$^{-1}$. These larger errors in this region may be a result of the model 
attempting to accurately fit both the high velocity red-component data points, which we 
credit to pure rotation, and the surrounding, lower velocity single-Gaussian data points, 
which are likely some combination of rotation and filament kinematic components.

With the red component of the double-peaked [SIII] lines being attributed to 
rotation, the remaining blue component velocities correspond to the inner portion 
of the filament. Thus, the full kinematic map of the filament in Mrk 509, as shown 
in Figure \ref{rotation}, can be seen by combining the 
blueshifted velocities of the single-peaked [SIII] lines west of the nucleus and 
the blue component of the double-peaked lines. In the inner portion of the filament, 
the `blue' component is actually close to systemic or slightly redshifted. These 
velocities can therefore be interpreted as the filament decelerating to systemic 
velocity as it reaches the corner of its check mark geometry and beginning to fall 
back toward the nucleus as redshifted velocities appear closer toward the nucleus. 
It is much more difficult to interpret the check mark as an outflow feature in the 
opposite direction because the filament would have to originate close to the nucleus, 
take a sharp right angle turn at $\sim$700 {\it pc}, and then accelerate outwards.

\section{Discussion and Conclusions}
\label{sec8}

\begin{figure}[h!]
\label{cartoon}
  \centering
  \includegraphics[width=0.45\textwidth]{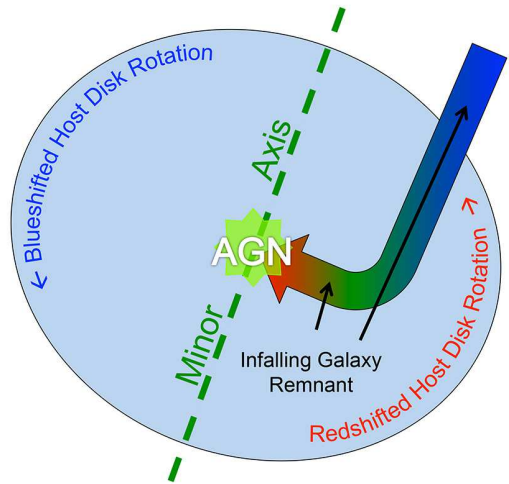}
  \caption{A cartoon interpretation of the IFU kinematics observed in Mrk 509. 
    Our measurements conform to a rotating gas disk with a 
    major axis near $PA=110^{\circ}$, except for the inflowing filament to 
    the northwest, which is blueshifted in the extended portion before 
    decelerating and turning toward the nucleus. It is unclear which side of the 
    host disk is closer, although either position does not affect our interpretation.}
\end{figure}

Measured velocities show kinematic components for {\it 1)} a rotating host disk and 
{\it 2)} the inner few kiloparsecs of the filament. We find that the linear portion 
of the filament, adjacent to the nucleus, contains blueshifted velocities that decelerate 
to systemic velocity as it nears the `elbow' of the check mark, and accelerate to 
redshifted velocities as the filament continues to travel toward the nucleus. Given the 
detailed kinematics now available for the filament in Mrk 509, in addition to its 
continuum emission, morphology, and orientation, we conclude that the filament structure 
cannot be related to typical NLR outflows \citep{Fis13,Fis14}. These kinematics instead 
suggest inflow where the filament is traveling toward us before decelerating as it 
interacts with the host disk, likely through dynamical friction \citep{Ken03}, and 
begins to fall back toward the nucleus, as shown in Figure \ref{cartoon}. 

In this 
two-component model, we do not account for redshifted velocities to the northeast of 
the nucleus (Figure \ref{ifu}, right). The source of these kinematics are unclear 
and not included in our analysis, as it cannot be determined whether they are inflowing 
or outflowing. Inflows may be present if the dwarf galaxy fell from high latitudes onto 
the host disk, crossed over the nucleus instead of falling into the nucleus, and then 
fell back toward the center where the red velocities are observed. Alternatively, these 
red velocities could be due to outflows, where the observed emission is one half of a 
biconical outflow and the symmetric blueshifted portion is hidden amongst the blueshifted 
velocities of the inflowing filament. 

Our current interpretation agrees with that given in \citet{Phi83}, where spectral 
analysis of low-ionization gas surrounding the nucleus conforms to a rotation pattern, 
and high-ionization gas in the same locations exhibit only blueshifted velocities. As 
we detect velocities of similar scale in both the rotation and highly ionized, inflowing 
components, we can attribute the difference in velocities between the low- and 
high-ionization gases seen in their results to the filament being bright in highly ionized 
line species (i.e. [O~III] and [S~III]) versus emission from the host disk. In measurements 
where emission lines are a combination of both the redshifted portion of the rotation 
curve and the bright, blueshifted filament, the peak of the emission lines would be 
shifted to shorter wavelengths. 

Although it is common to see the effects of interactions between galaxies as disturbances 
and tidal tails in their outer regions, minor mergers and the tidal striping of a dwarf 
galaxy as it falls towards the nucleus of a Seyfert galaxy are a rare occurrence \citep{Ken03}, 
as a only a small number of satellites will have an initial orbit path resulting in a direct 
merger onto the nucleus of the host galaxy versus simply being assimilated into the host galaxy 
disk. Mrk 509 embodies the first detected case of remains of an ongoing minor merger with a 
gas-rich dwarf galaxy. An explanation for the observed filament kinematics 
may then be that the linear, blueshifted portion is the orbital velocity of the 
merging dwarf galaxy before colliding with the inner host disk, and the inner redshifted 
portion shows pure gravitational infall into the nucleus after losing its orbital angular momentum.

\begin{figure}[h!]
\label{cont}
  \centering
  \includegraphics[width=0.45\textwidth]{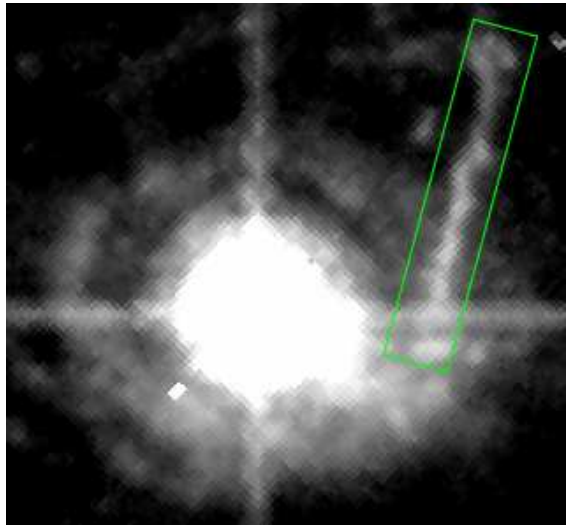}
  \caption{HST 550M continuum image of Mrk 509. Flux from the inflowing filament was measured in the boxed field and 
compared to the nucleus-subtracted flux of the host galaxy. From this comparison, the flux of the filament is $\sim1.5\%$ 
that of the host galaxy flux. As such, if the host galaxy of Mrk 509 is similar to the Milky Way, the filament is 
analagous to the Small Magellanic Cloud.}
\end{figure}

From available {\it HST} continuum imaging (Figure \ref{cont}), the measured flux of 
this filament corresponds to ∼1.5\% of the flux of the host galaxy of Mrk 509, a flux ratio similar to that between the 
Milky Way and the Small Magellanic Cloud. How efficient can we expect a minor merger 
such as this to be in fueling the host SMBH? Using M$_{BH} = 1.43 \times 10^{8} M_{\odot}$ 
\citep{Pet04} we calculate an Eddington luminosity of $L_{edd} = 1.8 \times 10^{46}$
erg s$^{-1}$. With an X-ray luminosity of $L_{2-10 keV} = 1.17 \times 10^{44}$ erg s$^{-1}$ 
(\citealt{Kra04}, and references therein) and a bolometric correction of 30 \citep{Awa01}, 
we calculate a bolometric luminosity of $L_{bol} = 3.5 \times 10^{45}$ erg s$^{-1}$, and 
an Eddington ratio of $L/L_{edd} = 0.19$ for Mrk 509. If we use the \citet{Kra04} [O~III] 
luminosity $L_{[OIII]} = 1.63 \times 10^{42}$ erg s$^{-1}$ instead of $L_{2-10 keV}$, we 
calculate the bolometric luminosity (using a correction of 3500; \citealt{Hec04}) to be 
$L_{bol} = 5.6 \times 10^{45}$ erg s$^{-1}$, with a resultant Eddington ratio of 
$L/L_{edd} = 0.31$. These values are high for typical Seyferts, which generally have 
$L_{bol}/L_{edd} < 0.1$ \citep{Ho09}. Thus, Mrk 509's high luminosity and high Eddington ratio 
are consistent with an ongoing fueling event. Measurements of additional infrared emission lines 
will allow us to determine gas densities and masses of the inflowing gas via photoionization models, 
allowing for a mass inflow measurement of the filamentary gas. Mass inflow and outflow rates will 
be revisited in a follow-up paper that includes a more thorough analysis of the emission lines in Mrk 509.

With limited observations available for Mrk 509, further study of this nearby AGN will improve 
our understanding of the feeding process in Seyfert galaxies and other AGN. One question 
that can be addressed in the near future is whether the starburst 
ring surrounding the AGN is connected to the merger event. Based on mid-IR lines, 
Mrk 509 has a 42\% starburst contribution to [NeII] 12.8$\mu$m line emission,
and a star formation rate of $\sim$8 M$_{sun}$ yr$^{-1}$ \citep{Mel08}. This signifies an ongoing 
young starburst, the age of which may correspond to the infall time of the companion. 
Previous numerical simulations of galaxy mergers by \citet{Cox08} concluded that 
primary hosts in high mass-ratio mergers, similar to the case observed in Mrk 509, 
are unlikely to experience increased star formation around their nuclei. As such, 
it is important to compare the stellar populations in the filament and 
starburst ring to determine whether or not they are related. 

Additional studies could also be performed on the extended filament outside 
of the NIFS field of view. Do the filament kinematics show signs of 
deceleration as it passes by the starburst ring? As the filament is decidedly 
brighter within the ring, as seen in the available {\it HST} imaging, this could 
signify an interaction between the filament and disk sooner than the elbow in the 
check mark morphology. Determining whether an early interaction takes place would 
be critical in interpreting the stellar populations, and the connection between the 
filament and starburst ring.

\acknowledgments
This material is based upon work supported by the National Science Foundation 
under Grant No. 1211651. This study was based on observations obtained at the Gemini Observatory 
(processed using the Gemini IRAF package), which is operated by the Association of 
Universities for Research in Astronomy, Inc., under a 
cooperative agreement with the NSF on behalf of the Gemini partnership: the National Science Foundation 
(United States), the National Research Council (Canada), CONICYT (Chile), the Australian Research 
Council (Australia), Minist\'{e}rio da Ci\^{e}ncia, Tecnologia e Inova\c{c}\~{a}o (Brazil) and 
Ministerio de Ciencia, Tecnolog\'{i}a e Innovaci\'{o}n Productiva (Argentina).
The authors would like to thank the referee for their helpful comments. T.C.F. would like to 
thank Rachel Kuzio de Naray for insightful conversations on disk kinematics and Crystal Pope for her role 
in assisting with kinematic measurements of the Mrk 509 NIFS observations. 
R.A.R thanks the support by CNPq (project n$^\circ$: 470090/2013-8) and FAPERGS (project n$^\circ$: 12/1209-6)

\bibliographystyle{apj}
\bibliography{apj-jour,mrk509_arxiv} 


\begin{figure*}
\label{ifu}
\centering
\begin{minipage}[h]{\textwidth}

  \begin{tabular}{ccc}


    \includegraphics[width=0.32\textwidth]{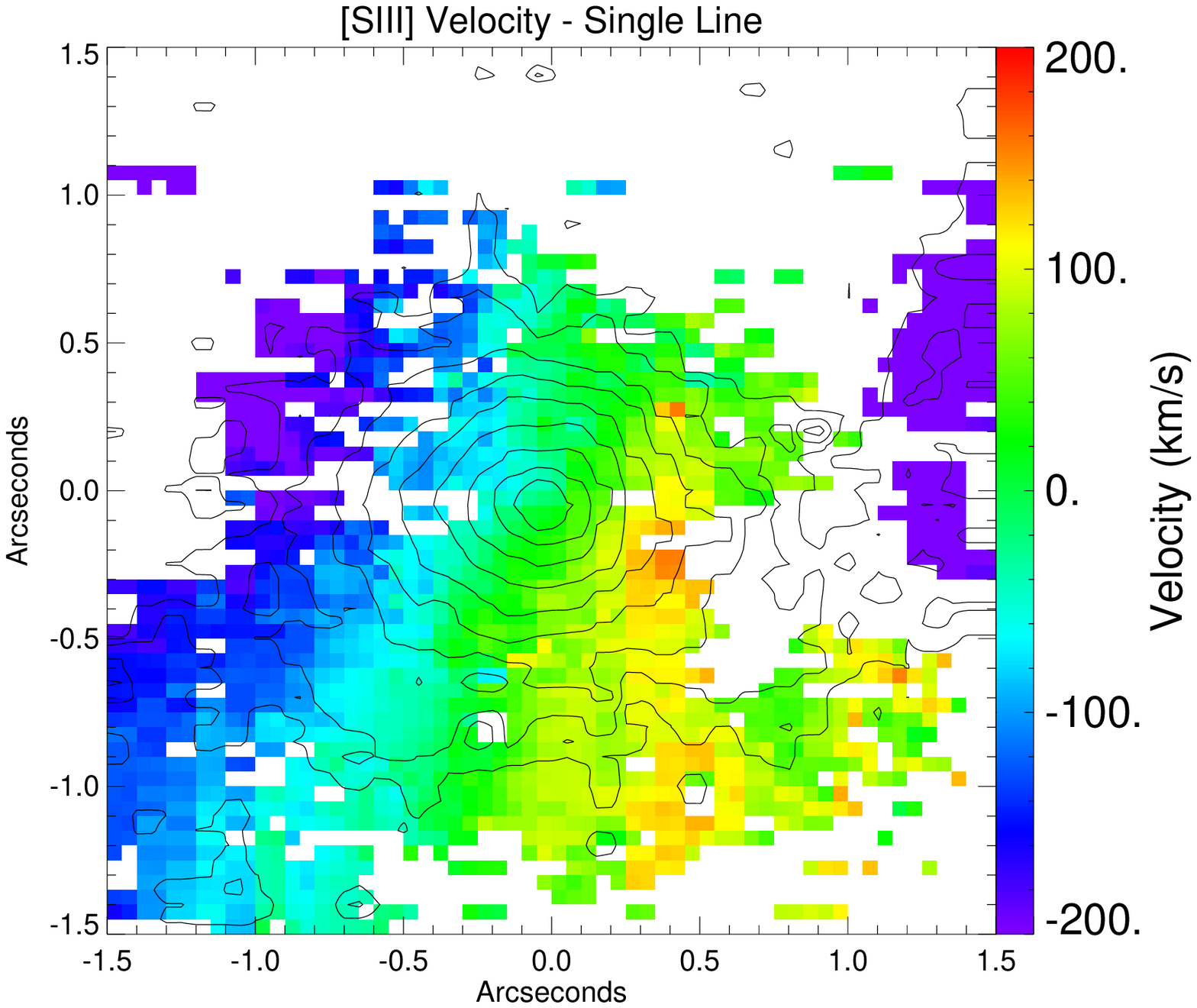}&

    \includegraphics[width=0.32\textwidth]{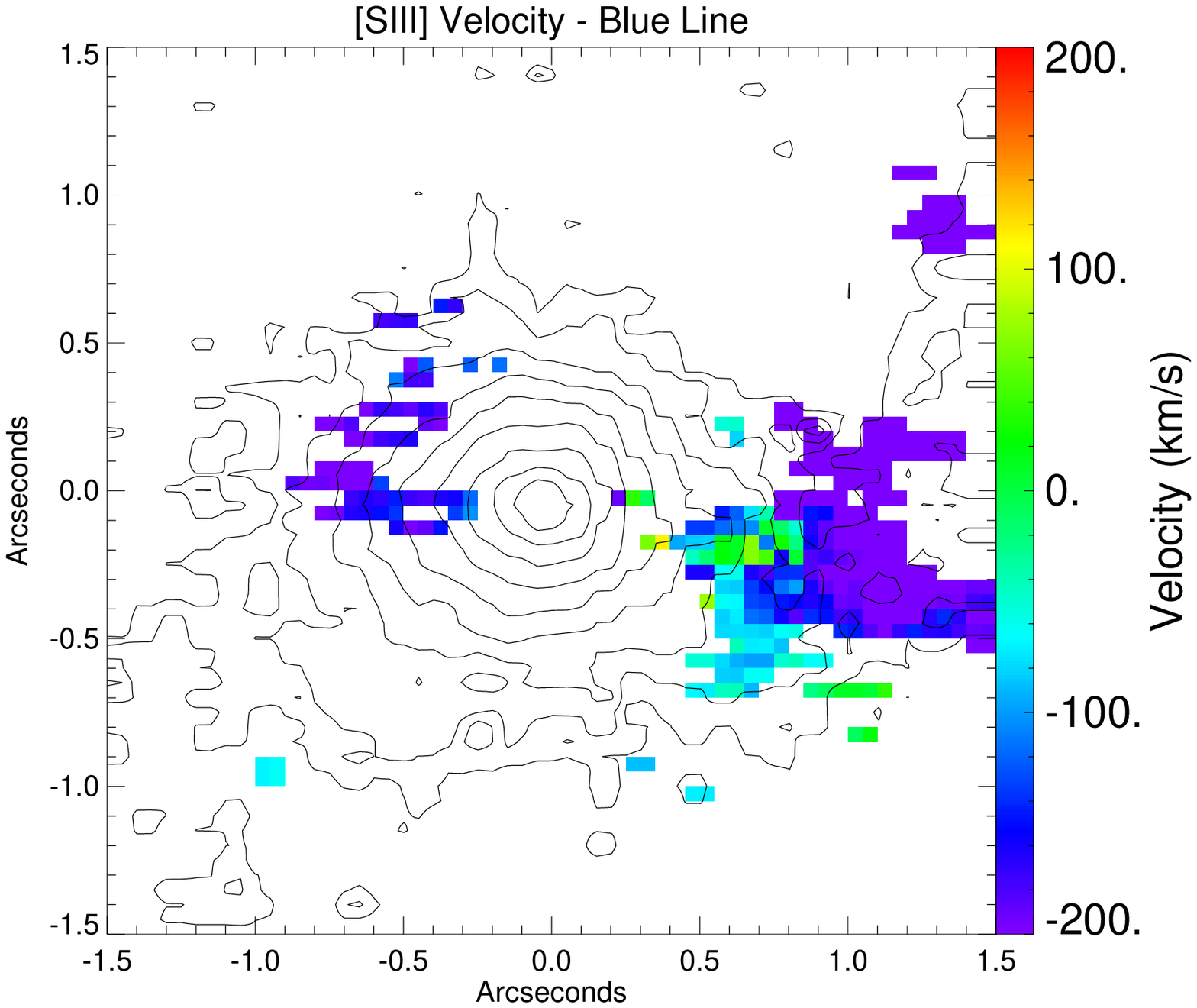}&

    \includegraphics[width=0.32\textwidth]{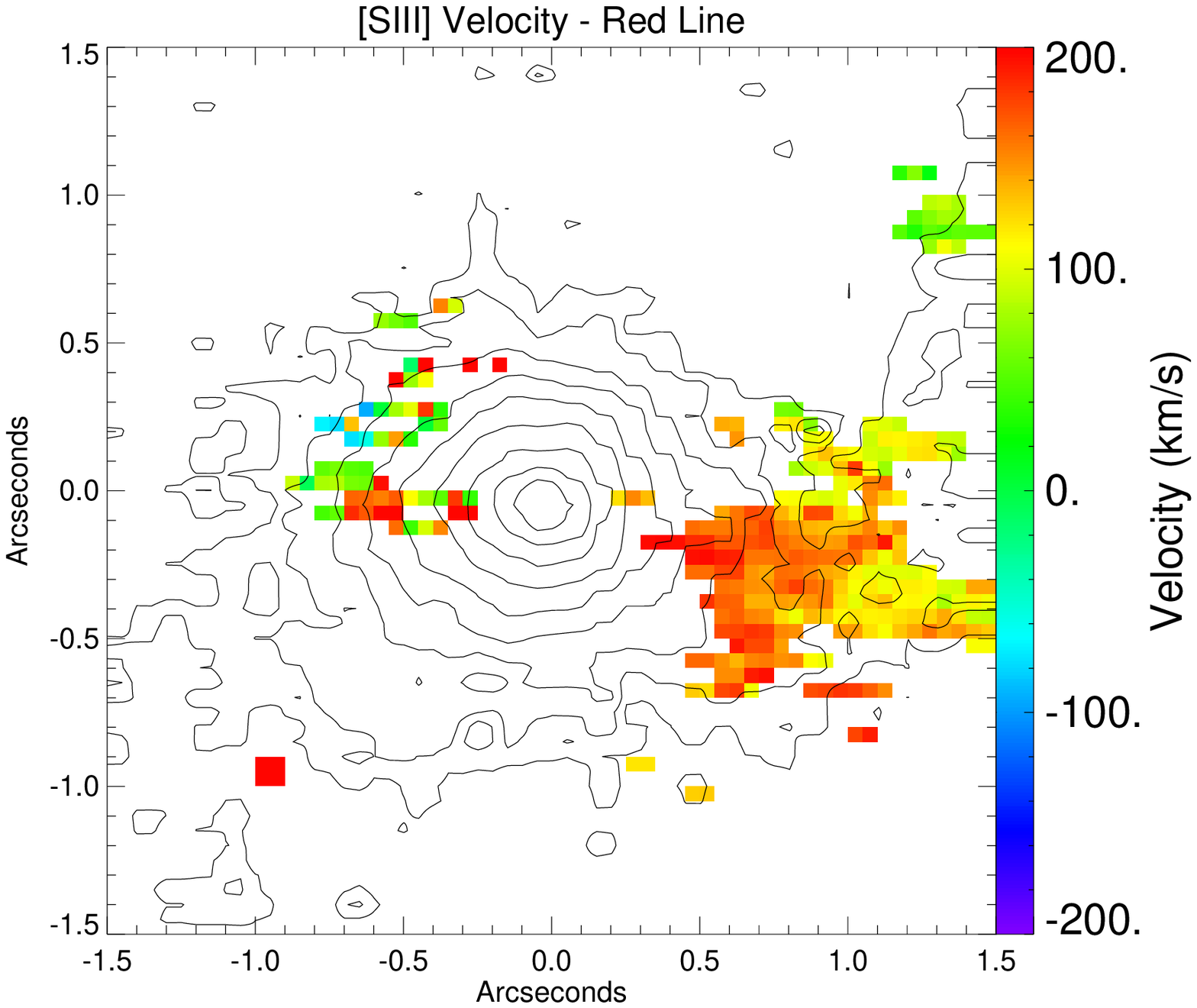}\\

    \includegraphics[width=0.32\textwidth]{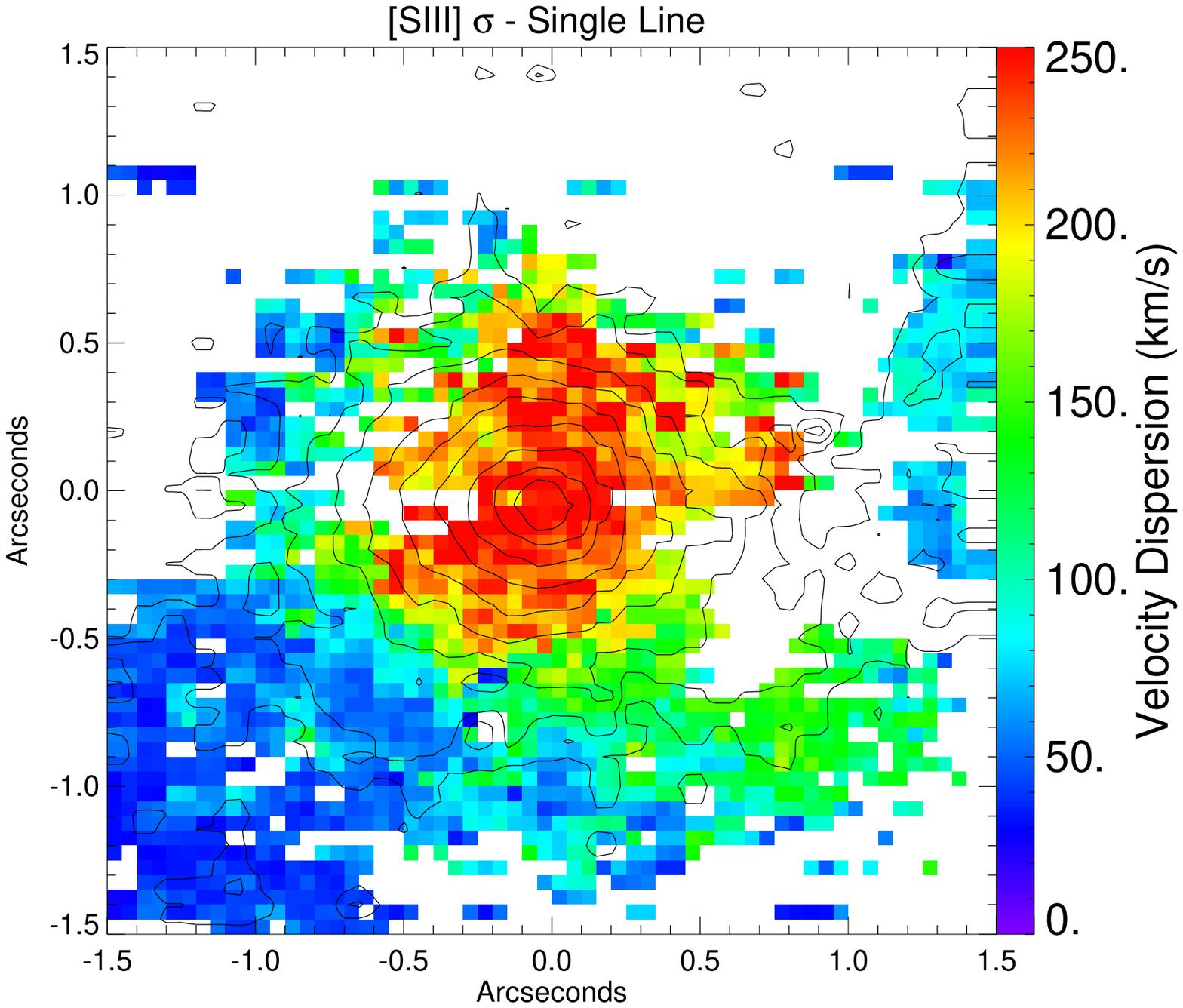}&

    \includegraphics[width=0.32\textwidth]{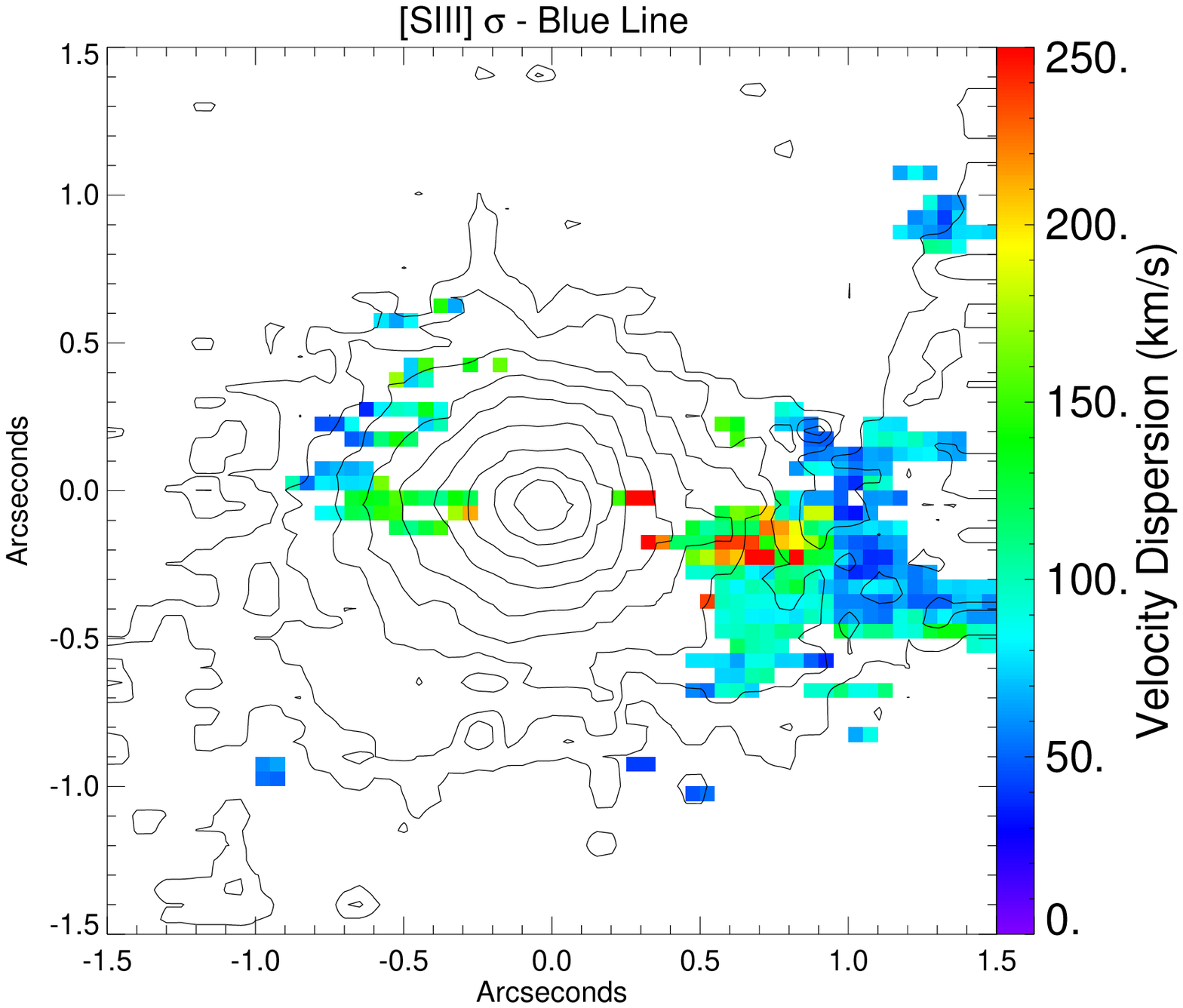}&

    \includegraphics[width=0.32\textwidth]{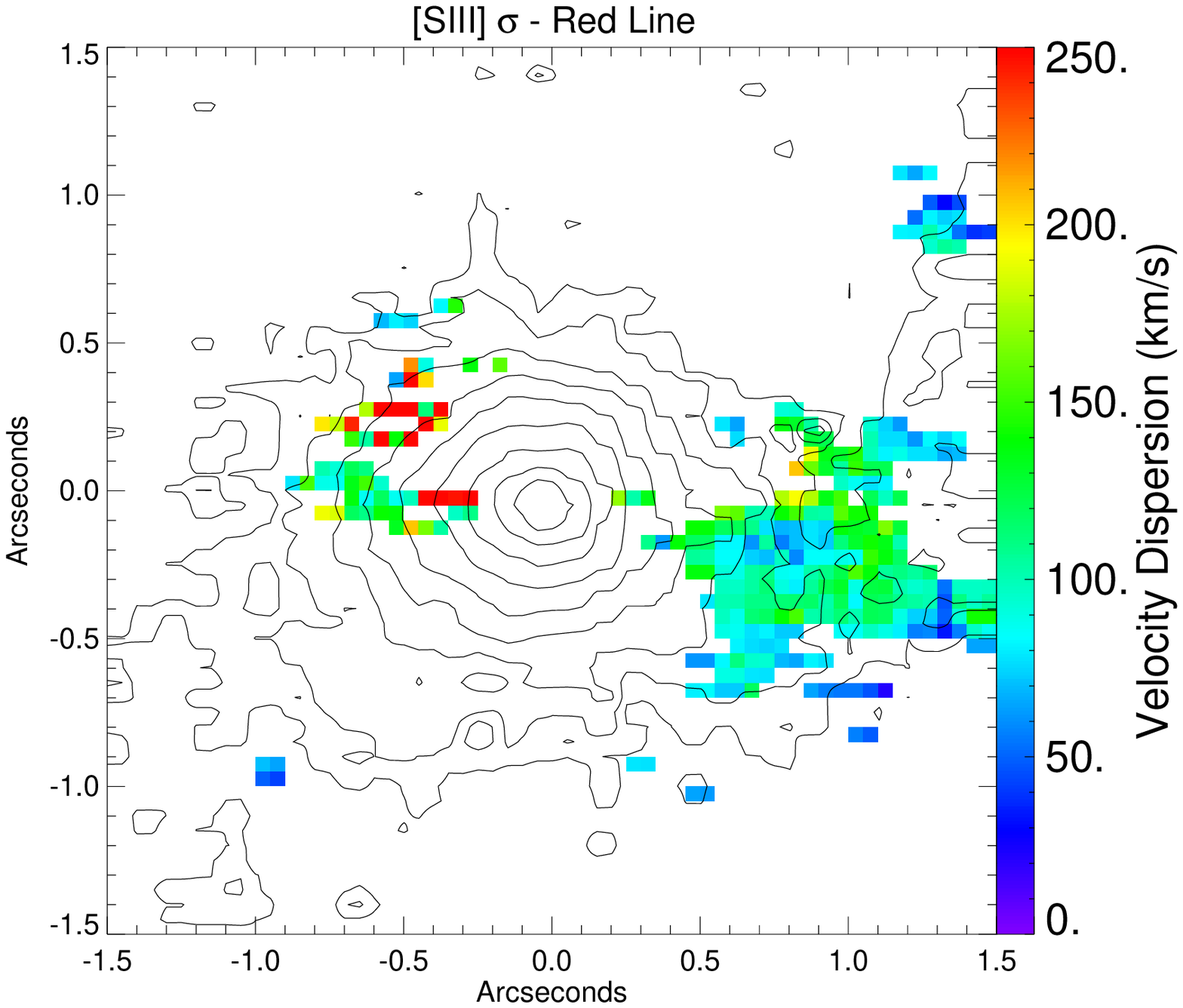}\\

  \end{tabular}
\caption{{\it Top:} Radial velocity maps for [S~III] emission successfully fit with a 
single Gaussian, the blue component of two Gaussians, and the red component of two Gaussians.
{\it Bottom:} Velocity dispersion maps for the same features. Continuum-subtracted [S~III] flux 
map contour lines are overlaid for all. North and east are at the top and left sides of the image 
respectively.}
\label{ifu}
\end{minipage}
\end{figure*}

\begin{figure*}
\label{rotation}

\centering
  \begin{tabular}{cc}


    \includegraphics[width=0.45\textwidth]{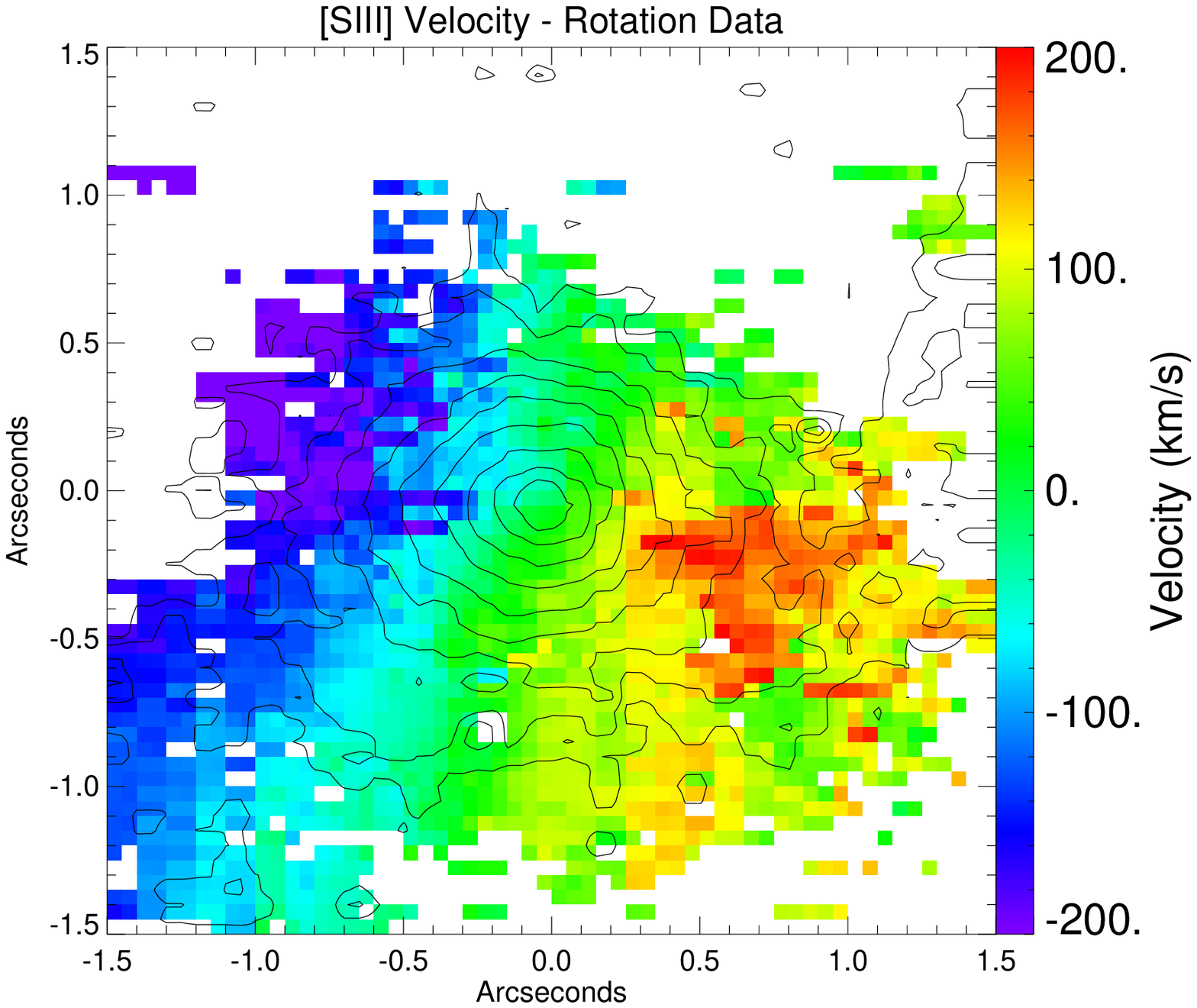}&
    
    \includegraphics[width=0.45\textwidth]{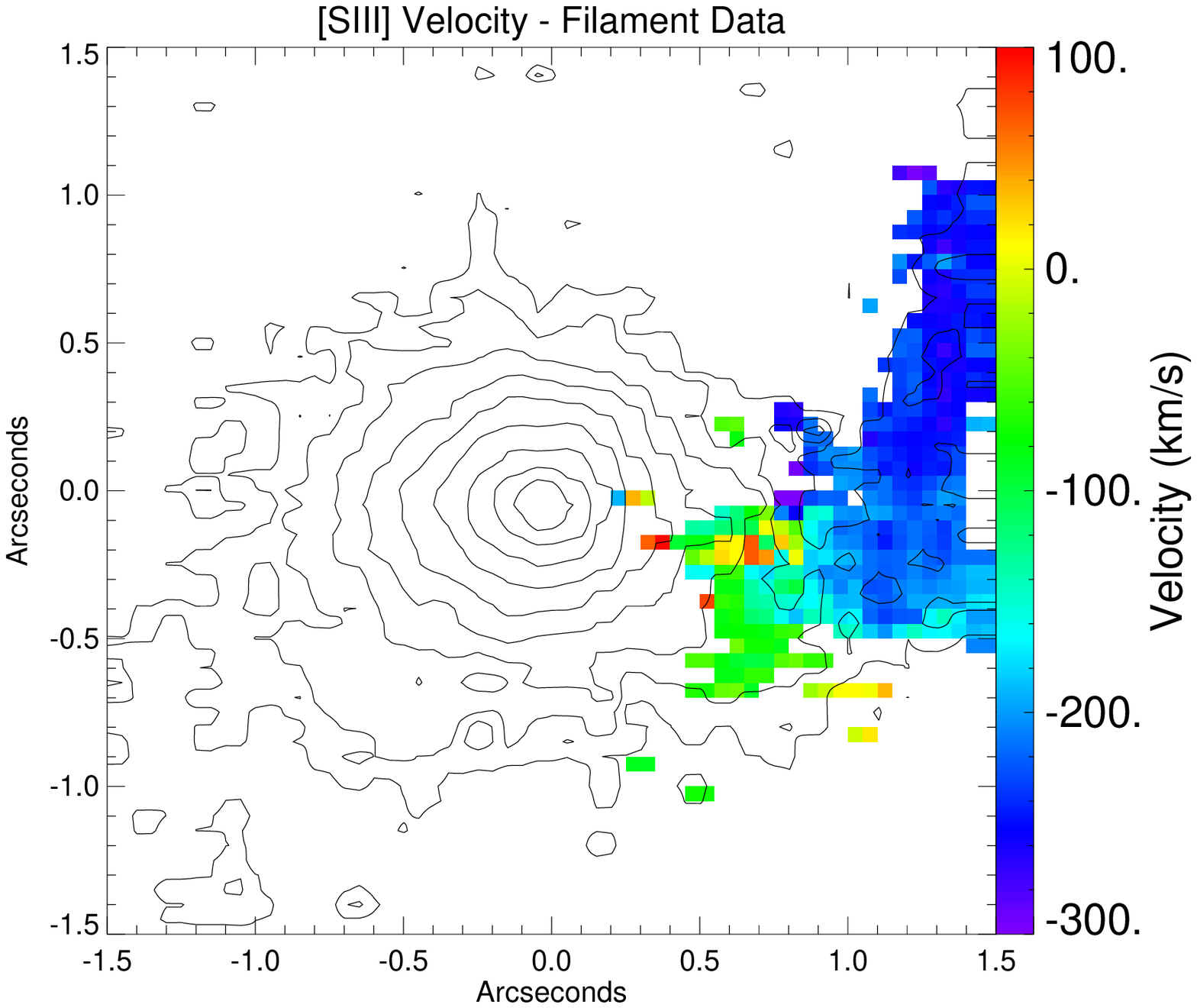}\\

    \includegraphics[width=0.45\textwidth]{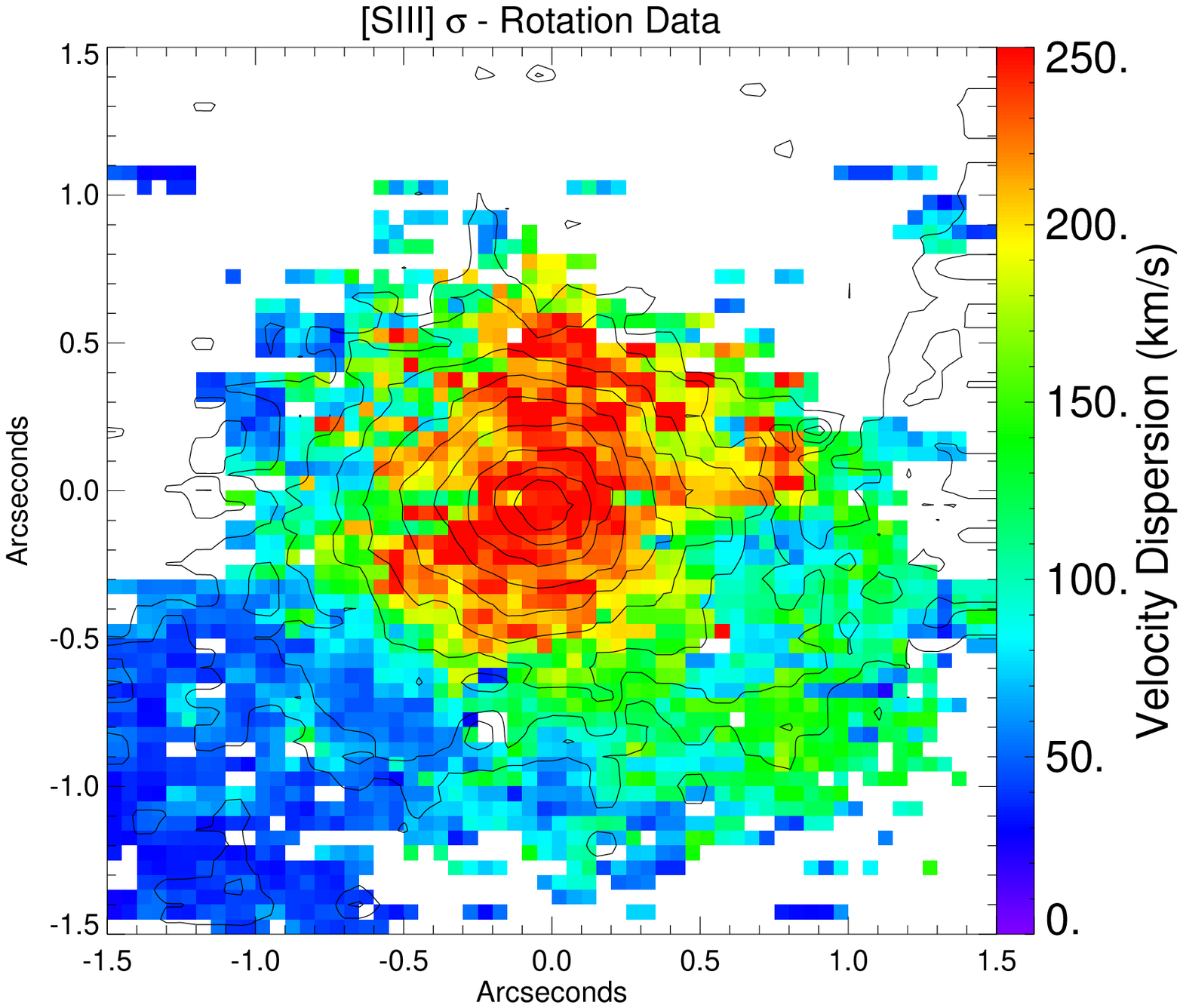}&

    \includegraphics[width=0.45\textwidth]{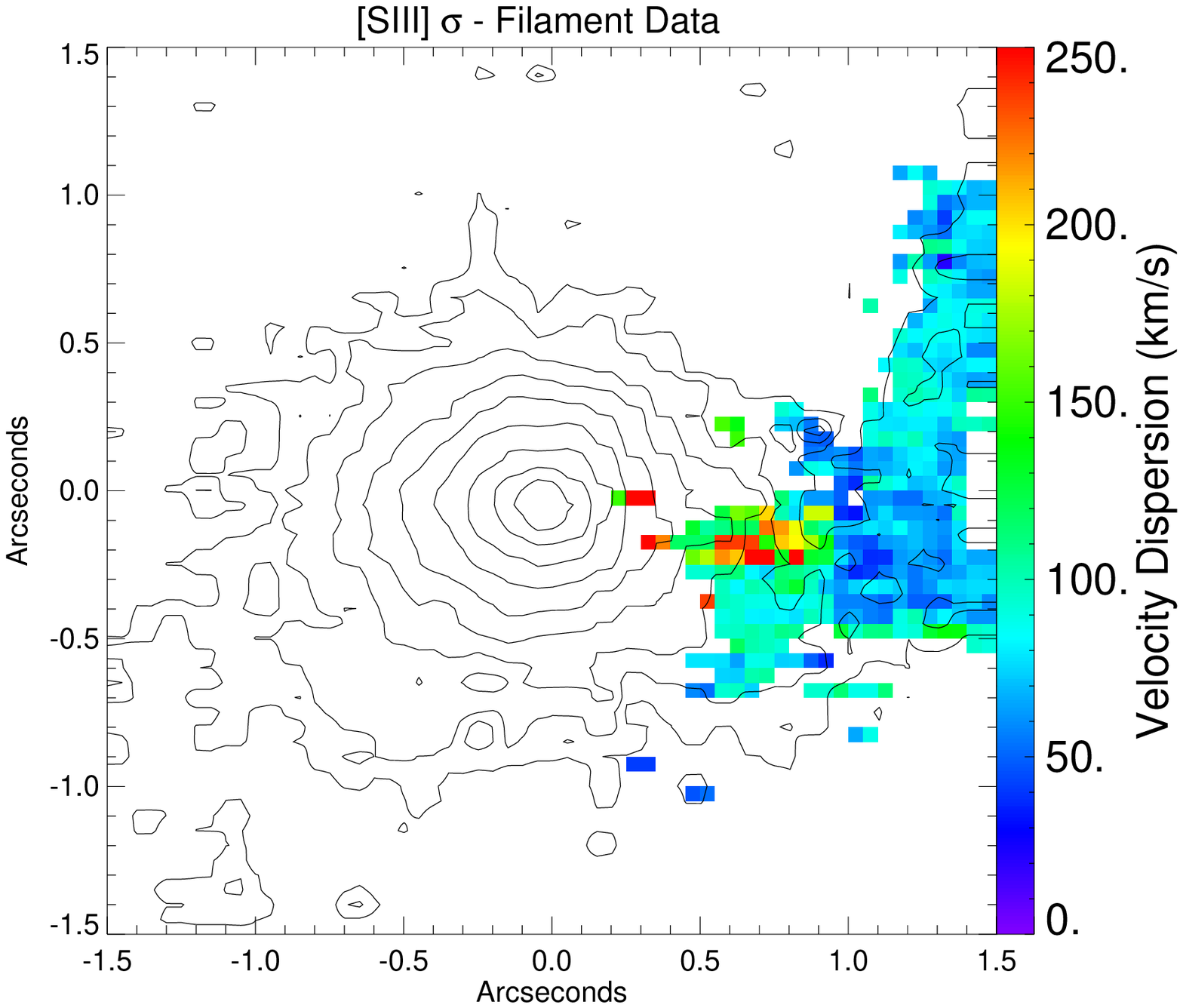}\\

  \end{tabular}
\caption{ {\it Upper Left:} Hypothesized rotation kinematics generated by combining 
kinematic measurements for the single-peaked [SIII] lines (minus northwest 
blueshifted velocities attributed to the filament) and the red component of the 
two Gaussian fit [SIII] lines. {\it Upper Right:} Filament kinematics generated by 
combining the remaining blueshifted velocities in the single-peaked [SIII] lines 
west of the nucleus and the blue component of the two Gaussian fit [SIII] lines.
{\it Lower Left:} Velocity dispersion map of rotation kinematics. {\it Lower Right:} 
Velocity dispersion map of filament kinematics. Continuum-subtracted [S~III] 
flux map contour lines are overlaid for all. North and east are at the top and left 
sides of the image respectively.}
\label{rotation}
\end{figure*}

\begin{figure*}
\label{model}

\centering
  \begin{tabular}{cc}

    \includegraphics[width=0.45\textwidth]{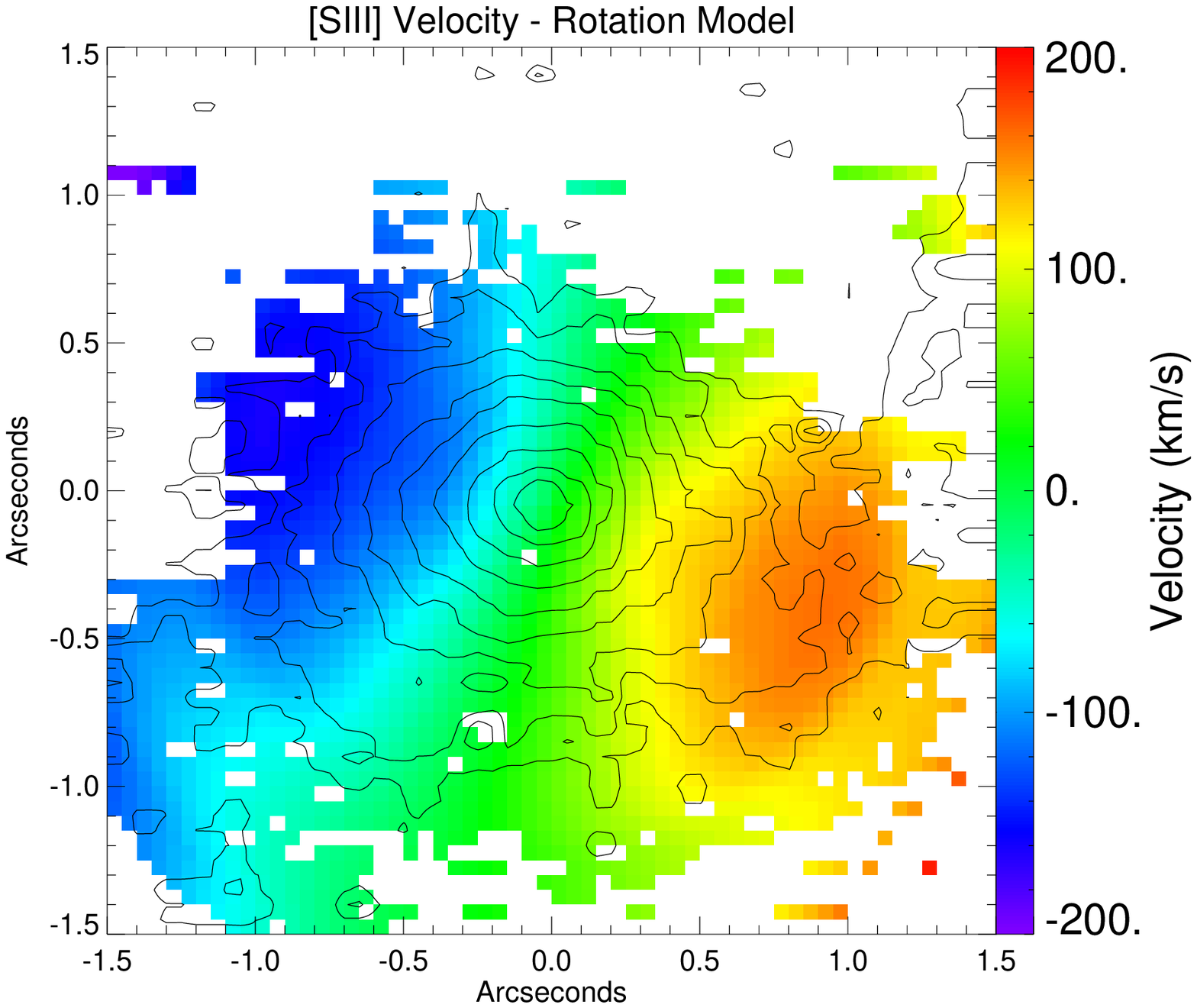}&

    \includegraphics[width=0.45\textwidth]{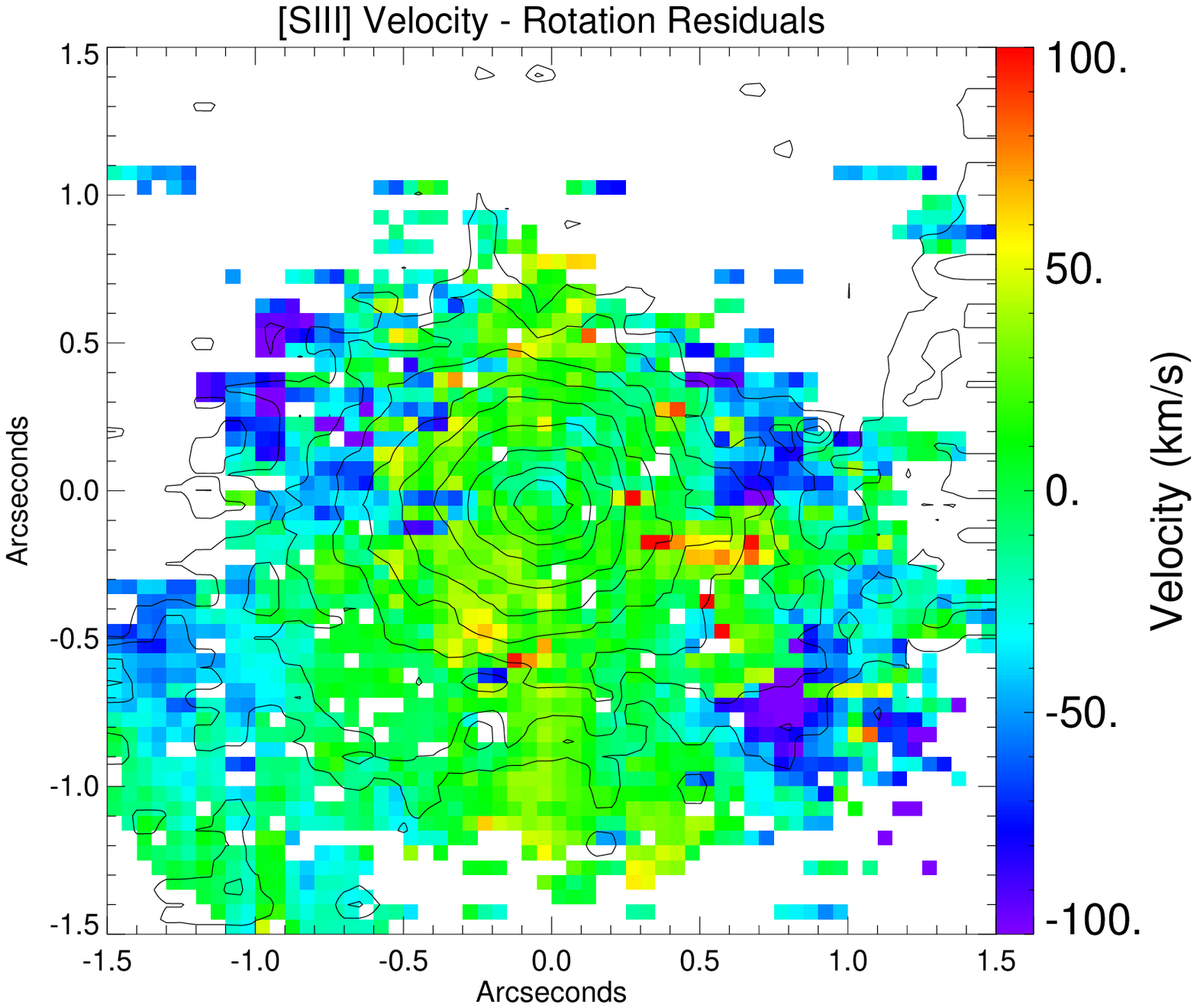}\\
  \end{tabular}
\caption{{\it Left:} Best fit rotating disk model using {\tt DiskFit}. 
  {\it Right:} Residual map for rotation data - rotation model. Continuum-subtracted [S~III] 
  flux map contour lines are overlaid for all. North and east are at the top and left 
  sides of the image respectively.}

\label{model}
\end{figure*}

\end{document}